\RequirePackage{fix-cm}
\documentclass[12pt]{article}
\usepackage{graphicx}
\usepackage{latexsym}
\usepackage{epsfig,amsmath,amssymb,amsfonts,bm,lineno}
\usepackage{subfig}
\usepackage{amstext}
\usepackage{array}
\usepackage{parskip}
\usepackage{color}

\usepackage{setspace}
\usepackage{latexsym}

\newcommand{\ds}{\displaystyle}
\newcommand{\beq}{\begin{equation}}
\newcommand{\eeq}{\end{equation}}
\newcommand{\beqq}{\begin{eqnarray}}
\newcommand{\eeqq}{\end{eqnarray}}
\newcommand{\p}{\partial}

\newcommand{\eps}{\varepsilon}
\newcommand{\x}{\mbox{\boldmath$x$}}

\begin{document}

\title{Narrow escape to small windows on a small ball modeling the viral entry into the cell nucleus}
\author{T. Lagache$^{1,2}$ and D. Holcman$^{1,3}$ \footnote{$^1$  Applied Mathematics and Computational Biology, Ecole Normale Sup\'erieure, France. $^2$ The NeuroTechnology Center at Columbia University Biological Sciences 901 NWC Building, 550 West 120th Street, New York, N.Y. 10027. tl2756@columbia.edu. $^3$ Newton Institute, Churchill college and DAMTP Cambridge CB30DS, United Kingdom.}}
\date{\today}

\maketitle
\begin{abstract}
A certain class of viruses replicates inside a cell if they can enter the nucleus through one of many small target pores, before being permanently trapped or degraded. We adopt for viral motion a switching stochastic process model and we estimate here the probability and the conditional mean first passage time for a viral particle to attain alive the nucleus. The cell nucleus is covered with thousands of small absorbing nuclear pores and the minimum distance between them defines the smallest spatial scale that limits the efficiency of stochastic simulations. Using the Neuman-Green's function method to solve the steady-state Fokker-Planck equation, we derive asymptotic formula for the probability and mean arrival time to a small window for various pores' distributions, that agree with stochastic simulations. These formulas reveal how key geometrical parameters defines the cytoplasmic stage of viral infection.
\end{abstract}
\section{Introduction}
How particles such as molecules, proteins, DNA, RNA, or viruses are moving inside the complex and crowded cellular environment \cite{Medalia} remains a challenge both experimentally  and theoretically.  For example, vesicles or RNA granules \cite{bressloff1} have to reach small targets in order to deliver their payload or trigger protein synthesis. However, large DNA or plasmid are too large and cannot pass the cytoplasmic crowded organization \cite{Dauty}. In some cases, large particles are transported intermittently along microtubules (MTs) toward the nucleus and Brownian motion inside the cytoplasm. Many viruses containing DNA have the ability to hijack the cellular transport machinery to reach a nuclear pore and deliver their genetic material inside the nucleus \cite{sodeik,Greber}. Although viral trajectories can be monitored \textit{in vivo} using live microscopy, for viruses such as HIV or the Adeno-Associated Virus  \cite{charneau,Seisenberger}, these trajectories consist of alternating epochs of diffusion and directed motion. The precise nature of these trajectories remains unclear. In addition, on their way to the nucleus, viruses can be trapped in
the cytoplasm or degraded through several pathways (including the ubiquitin-proteasome).

\begin{figure}[http!]
\includegraphics[width=7.5cm]{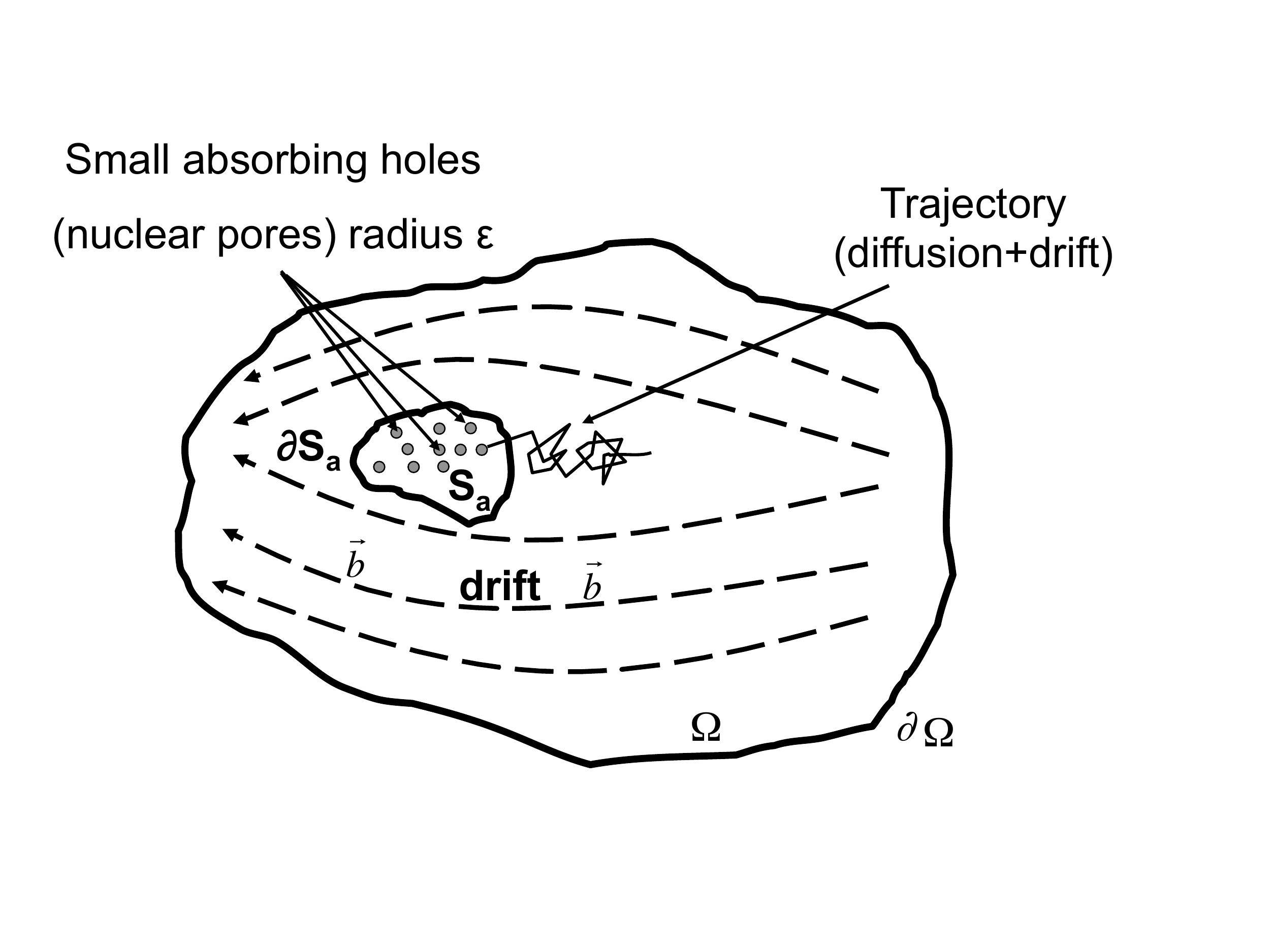}
\includegraphics[width=7.5cm]{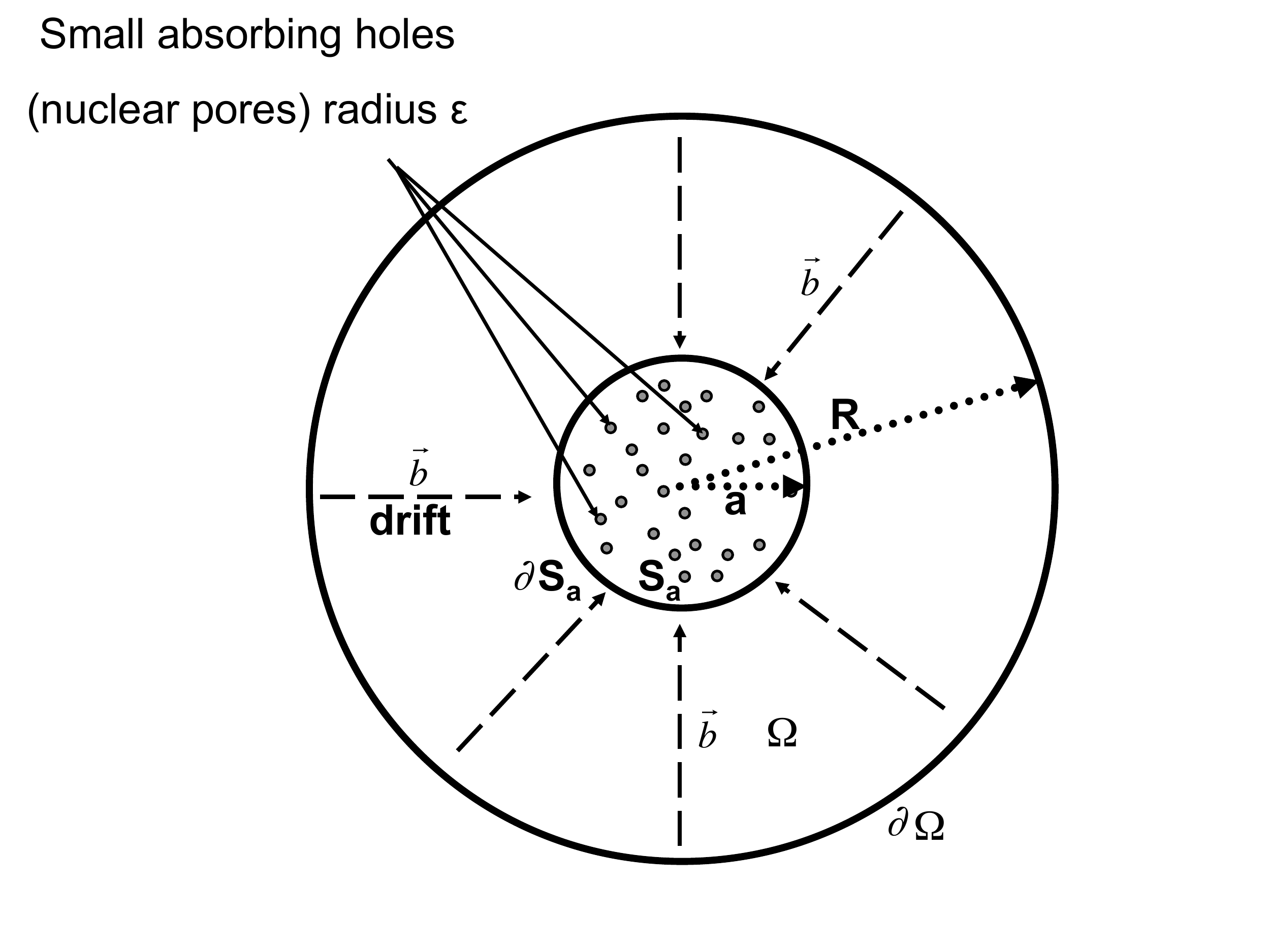}
\caption{\small{{\bf Schematic representation of the cell cytoplasm as a 3-dimensional domain
$\Omega$.} (Right-hand side): Stochastic trajectories, solutions of eq. \ref{langevin}, that contains both a diffusion and a drift terms can be absorbed at small windows with radius $\epsilon \ll |\Omega|^{1/3}$ located on the surface of the nucleus $\p S_a$. Right-hand side: simplified
spherical cell (radius $R$) containing a spherical nucleus that we model with a ball of radius $a$ such that $\epsilon \ll a\ll|\Omega|^{1/3}$.\label{FIGURE1}}}
\end{figure}

To quantify the success of the early steps of viral infection, we recently used a modeling approach at the single particle level \cite{david,PRE-manu}. Due to small size of the nuclear pore (Fig. \ref{FIGURE1}), Brownian simulations are always ineffective to estimate precisely the moments associated to arrival time. To study the dependency with respect to geometrical and dynamical parameters, we analyzed the intermittent stochastic dynamics of viruses along MTs and derive asymptotic formula for the conditional mean first passage time (MFPT) $\tau_n$ and the probability $P_n$ that a single particle arrives to $n$ small targets \cite{david,PRE-LH}. However, our previous formula are valid when the number of absorbing holes (targets) is not too large. We extend here our analysis to the case of many holes. This analysis relies on the explicit expansion of the Neumann-Green's function to order three \cite{Ward2}. The present method is also valid for many interacting small
holes \cite{Holcman-schussJPhysA,Holcman-schussPLA,Cheviakov}.

The paper is organized as follows. We first recall our previous model of viral particles and the stochastic description of trajectories. Second, we extend the small hole interaction method to the case of a stochastic particle with a drift and derive the mean arrival time using an interaction matrix between holes. Using the precise expansion of the
Neumann-Green's function for the sphere, we will derive new asymptotic formula for the probability and the mean conditioning time to reach one of the many nuclear pore. Finally, we confirm our analysis with some Brownian simulations. The new formula that significantly improve our previous effort \cite{david,PRE-LH} can now be used to study more precisely the first steps of viral infection in cells.

\section{The mean time to a small nuclear pore.}
Intermittent trajectories of a viral particle $\mathbf{x}(t)$ were described by the switching stochastic rule \cite{LagacheSIAM2008}
\beq
d\mathbf{x}= \left\{\begin{array}{l} \sqrt{2D}
d\mathbf{w} \quad\mbox{ when }\quad \mathbf{x}\left(t\right) \quad\mbox{ is free }\\ \\
{\bf V}dt \quad\mbox{ \, \, \, \, \, }\quad \mathbf{x}\left(t\right)
\quad\mbox{ bound },
\end{array}\right.\label{eq1}
\eeq
where $\bf w$ is a standard 3d-Brownian motion, $D$ the diffusion
constant and $\bf V$ the velocity of the directed motion along MTs
(randomly distributed). The switching dynamics depends on the
attachment and detachment rates \cite{LagacheSIAM2008}. We coarse-grained this switching
process into by a steady-state stochastic equation
\beq
d\mathbf{x} = \mathbf{b(x)}dt+\sqrt{2D}d \bf w,
 \label{langevin}
 \eeq
where the effective drift $\mathbf{b(x)}$ was found using the following criteria: inside the cytoplasm $\Omega$, the mean first passage time of stochastic processes \ref{eq1} and \ref{langevin} is the same \cite{LagacheSIAM2008,PRE-LH}. The drift $\mathbf{b(x)}$ depends on the cell geometry, the number and distribution of MTs and the rates of binding and unbinding of the particle to MTs.

Most viruses have to reach one of the small circular pore, modeled as absorbing windows of radius $\epsilon\ll1$ located on the boundary $\p S_a$ of the nucleus. We approximate a small pore as a small sphere $S_a$.  The external cell membrane defines the boundary $\p \Omega$ for the stochastic process eq. \ref{langevin}. The cell cytoplasm is represented as the three-dimensional bounded domain $\Omega$, whose boundary is $\partial \Omega \bigcup \p S_a$. It consists of a reflecting part except for the $n-$small absorbing windows $\partial N_a$ located on the nucleus
(fig.\ref{FIGURE1}-left).

Finally, we model the degradation activity in the cytoplasm by a steady-state killing rate $k(\bf{x})$ and a trajectory described by eq. \ref{langevin} can thus disappear before reaching the absorbing boundary $\partial N_a$ . The survival probability density function (SPDF) is solution of the forward Fokker-Planck equation \cite{HMS}
\beq
\begin{array}{c}
 \ds{\frac{\partial p}{\partial t}}=\Delta p-\nabla \cdot \mathbf{b}p-kp  \label{SPDF}\\
 p\left( {\mathbf{x},0} \right)=p_i \left(\mathbf{x} \right) \\
 \end{array}
\eeq
with the boundary conditions:
\beq
 p(\mathbf{x},t)=0 \hbox{ on } \p N_a \hbox{ and }
\mathbf{J}(\mathbf{x},t).\mathbf{n}_{\mathbf{x}} = 0 \hbox{ on }
\partial \Omega \bigcup \left(\p S_a - \p N_a\right)
\eeq
where the flux density vector is
\beq
\mathbf{J}(\mathbf{x},t)=-D \nabla p(\mathbf{x},t)
+\mathbf{b(x)}p(\mathbf{x},t).\label{Ji}
 \eeq
where $\mathbf{n}_{\mathbf{x}}$ is the normal derivative at a point $\mathbf{x}$.

We recall that the mean probability $\langle P \rangle $ and the conditional MFPT $\langle \tau \rangle$ (averaged over the initial particle distribution) for a stochastic particle driven by eq. \ref{langevin} to
reach the boundary $\p N_a$ before degradation can be expressed using  $\tilde{p}(\mathbf{x})=\int_0^{\infty} p(\mathbf{x},t) dt$ and $q(\mathbf{x})=\int_0^{\infty} t p(\mathbf{x},t) dt$ \cite{david} as
\beq
\ds{\langle P \rangle}(n,\epsilon) =\ds{1-\int_{\Omega}k(\mathbf{x})\tilde{p}(\mathbf{x})
d\mathbf{x}},\label{PN}
\eeq
and
\beq
\ds{\langle \tau \rangle}(n,\epsilon)=\frac{\ds{\int_{\Omega} \tilde{p}(\mathbf{x})
d\mathbf{x}-\int_{\Omega} k(\mathbf{x}) q(\mathbf{x})
d\mathbf{x}}}{\ds{1-\int_{\Omega}k(\mathbf{x})\tilde{p}(\mathbf{x})
d\mathbf{x}}}.
\label{TN}
\eeq
For a potential drift $\mathbf{b\left( x \right)}=-\nabla \Phi \left( \mathbf{x} \right)$, an asymptotic expansion in the small parameter $\epsilon$ \cite{PRE-LH}, reveals that
\beq
\left\{\begin{array}{l} \langle P \rangle(n,\epsilon)= \frac{\ds{e^{-\frac{\Phi_0}{D}}}}
{\ds{\frac{1}{4 D
n\epsilon}\int_{\Omega}e^{-\frac{\Phi(\mathbf{x})}{D}}k(\mathbf{x})d\mathbf{x}+e^{-\frac{\Phi_0}{D}}}}
\\ \\
\langle \tau \rangle(n,\epsilon)=\frac{\ds{\frac{1}{4D n \epsilon}\int_{\Omega}e^{-\frac{\Phi(\mathbf{x})}{D}}d\mathbf{x}}}
{\ds{\frac{1}{4 D n
\epsilon}\int_{\Omega}e^{-\frac{\Phi(\mathbf{x})}{D}}k(\mathbf{x})d\mathbf{x}+e^{-\frac{\Phi_0}{D}}}},
\end{array}\right. \label{former_formula}
\eeq
where $\Phi_0$ is the constant value of the radial potential $\Phi(\mathbf{x})$ on the centered nucleus where the nuclear pores are uniformly distributed. The range of validity of these asymptotic expressions has been explored with Brownian simulations for a single hole \cite{PRE-manu}. However, these formulas do not account for the possible interactions between the small absorbing pores, and for a large number of nuclear pores $n\gg 1$,
\beq
\lim_{n \rightarrow \infty, n\epsilon^2\ll1}\langle \tau \rangle (n,\epsilon)=0,
\eeq
which shows the limitation of the previous formula.

We find here the correction term that accounts for the nuclear geometry. Interactions between absorbing windows can
drastically affect the MFPT \cite{Holcman-schussPLA,Ward1}, and we study here these interactions and extend the narrow escape time for a stochastic particle (with a drift) in the presence of a killing field $k(\x)$ to reach one of the interacting absorbing windows located on the nucleus.  We obtain an estimate for the probability $\langle P \rangle$ and the associated conditional MFPT $\langle \tau \rangle$. Both quantities are solutions of a coupled system of partial differential equations. For a large number of holes covering homogeneously the nucleus, we extend our analysis using a mean field approximation and obtain formulas for the probability $\langle P \rangle$ and the mean time $\langle \tau \rangle$, valid for a large range of both parameters $\epsilon$ and $n$, generalizing formula \ref{former_formula}. Finally, we test the asymptotical results against Brownian simulations and apply our formula to model viral trafficking ($\mathbf{b\neq 0}$) and non-viral gene vectors (Brownian diffusion $\mathbf{b=0}$) that have to reach one of the $n\approx 2,000 \gg 1$ \cite{maul} nuclear pores covering the nucleus in order to deliver their genetic material inside the cell nucleus.

\section{Asymptotic derivations of the mean time  $\langle \tau \rangle$ and the probability $\langle P \rangle$}
The $n-$absorbing windows $ \p N_a = \bigcup_{i=1}^{n}
\p \Omega_i$ have the same radius $\epsilon$, centered at positions
$\left(\mathbf{x}_i\right)_{i=1}^{n}$. The steady state SPDF $p$ is solution of eq. \ref {SPDF}) \cite{HMS}.

The Neumann-Green function $\mathcal{N}(\mathbf{x},\mathbf{x}_0)$ is solution of the differential
equation \cite{david}
\beqq
\Delta \mathcal{N} (\mathbf{x}, \mathbf{x}_0)&=&-\delta_{\mathbf{x}_0}(\mathbf{x}), \hbox{ }\mathbf{x} \in
\Omega,\nonumber \\
D \frac{\p \mathcal{N}}{\p n}
(\mathbf{x},\mathbf{x}_0)&=&-\frac{1}{|\p \Omega|} \hbox{
}\mathbf{x} \in \p \Omega.\label{Neumann}
\eeqq
We recall that $\tilde{p}(\mathbf{x})=\int_0^{\infty} p(\mathbf{x},t)dt$ is solution of equation
\beq \label{ptilde}
 \Delta \tilde{p}-\nabla \cdot (\mathbf{b}\tilde{p})-k\tilde{p}=-p_i
 \eeq
with the boundary conditions
\beq
\tilde{p}(\mathbf{x})=0 \hbox{ on } \p N_a=\bigcup_{i=1}^{n} \p \Omega_i \hbox{ and } \mathbf{\tilde{J}}(\mathbf{x}).\mathbf{n}_{\mathbf{x}} = 0 \hbox{ on } \partial \Omega \bigcup \left(\p S_a - \p N_a\right) \eeq
where $\mathbf{\tilde{J}}(\mathbf{x})=-D \nabla \tilde{p}(\mathbf{x}) +\mathbf{b(x)}\tilde{p}(\mathbf{x})$.

Green's identity gives
\beqq
I&=&\int_{\Omega}\left(\Delta \tilde{p}(\mathbf{x})-\nabla \cdot
\mathbf{b}\tilde{p}(\mathbf{x})-k\tilde{p}(\mathbf{x})\right)\mathcal{N}(\mathbf{x},\mathbf{x}_0)d\mathbf{x}\nonumber
\\&-&\int_{\Omega}\Delta \mathcal{N}(\mathbf{x},\mathbf{x}_0)
\tilde{p}(\mathbf{x})d\mathbf{x}.
\eeqq
Consequently, we have
\beq
I=-\int_{\Omega}p_i(\mathbf{x})
\mathcal{N}(\mathbf{x},\mathbf{x}_0)+\tilde{p}(\mathbf{x}_0)
\eeq
and from Green's identity
\beqq
I&=&-\int_{\partial N_a}
\mathbf{\tilde{J}}(\mathbf{x}).\mathbf{n}_{\mathbf{x}}\mathcal{N}(\mathbf{x},\mathbf{x}_0)
d\mathbf{x}+ \int_{\Omega} \mathbf{b}(\mathbf{x}).\nabla
\mathcal{N}(\mathbf{x},\mathbf{x}_0) \tilde{p}(\mathbf{x})
d\mathbf{x} \nonumber\\&-&\int_{\Omega}k(\mathbf{x})
\tilde{p}(\mathbf{x}) d\mathbf{x}+\frac{1}{|\partial
\Omega|}\int_{\p \Omega}\tilde{p}(\mathbf{x})d\mathbf{x}.
\eeqq
Thus, we obtain
\beqq
 \int_{\Omega}\left(k(\mathbf{x})
\tilde{p}-p_i(\mathbf{x})\right)
\mathcal{N}(\mathbf{x},\mathbf{x}_0)d\mathbf{x}&=&-\int_{\partial
N_a}
\mathbf{\tilde{J}}(\mathbf{x}).\mathbf{n}_{\mathbf{x}}\mathcal{N}(\mathbf{x},\mathbf{x}_0)
d\mathbf{x}\nonumber\\&+&\int_{\Omega} \mathbf{b}(\mathbf{x}).\nabla
\mathcal{N}(\mathbf{x},\mathbf{x}_0) \tilde{p}(\mathbf{x})
d\mathbf{x}\nonumber \\&+&\frac{1}{|\partial \Omega|}\int_{\p
\Omega}\tilde{p}(\mathbf{x})d\mathbf{x}-\tilde{p}(\mathbf{x}_0).
\label{I}
\eeqq
When the field is the gradient of a potential and when the first eigenvalue only contribute to the spectrum, the solution $\tilde{p}(\mathbf{x})$ is the steady-state, thus:

%
\beq
\tilde{p}(\mathbf{x})\approx C_{\epsilon}e^{-\frac{\Phi(\mathbf{x})}{D}}+O\left(1\right)\label{LTAP}.
\eeq
Furthermore,
\beq
q(\mathbf{x})=\left(C_{\epsilon}^2 \int_{\Omega}e^{-\frac{ \phi(\mathbf{x})}{D}}d\x\right)e^{-\frac{\Phi(\mathbf{x})}{D}}+O\left(1\right) \label{LTAQ},
\eeq
For a smooth initial distributions $p_i$, the integral
\beq
\int_{\Omega}p_i(\mathbf{x})\mathcal{N}(\mathbf{x},\mathbf{x}_i)d\mathbf{x}
\eeq
is uniformly bounded as $\epsilon \rightarrow 0$, while all other terms in relation \ref{I} are unbounded. Consequently, for a small degradation rate $k \ll 1$ limit, the integral equation (\ref{I}) is to leading order:
\beqq
\tilde{p}(\mathbf{x}_0)+O\left(1\right)&=&-\int_{\partial N_a}
\mathbf{\tilde{J}}(\mathbf{x}).\mathbf{n}_{\mathbf{x}}\mathcal{N}(\mathbf{x},\mathbf{x}_0)
d\mathbf{x}+\int_{\Omega} \mathbf{b}(\mathbf{x}).\nabla
\mathcal{N}(\mathbf{x},\mathbf{x}_0) \tilde{p}(\mathbf{x})
d\mathbf{x}\nonumber \\&+&\frac{1}{|\partial \Omega|}\int_{\p
\Omega}\tilde{p}(\mathbf{x})d\mathbf{x}.
\label{I2}
\eeqq
For $\x_0$ at a distance $O\left(1\right)$ away from absorbing
windows, $\mathcal{N}(\mathbf{x},\mathbf{x}_0)$ is uniformly bounded
for $\x
\in \p \Omega_a$. In addition, integrating (\ref{ptilde}) over
$\Omega$ we obtain:
\beq
\int_{\partial N_a}
\mathbf{\tilde{J}}(\mathbf{x}).\mathbf{n}_{\mathbf{x}}
d\x=1-\int_{\Omega}k(\mathbf{x})\tilde{p}(\mathbf{x})d\mathbf{x}=\langle
P \rangle \in [0,1].\label{conservation}
\eeq
Consequently, for $\x_0$ at a distance
$O\left(1\right)$ away from absorbing windows,
$\int_{\partial N_a}
\mathbf{\tilde{J}}(\mathbf{x}).\mathbf{n}_{\mathbf{x}}\mathcal{N}(\mathbf{x},\mathbf{x}_0)
d\mathbf{x}$ is uniformly bounded, and
\beq
\frac{1}{|\partial
\Omega|}\int_{\p
\Omega}\tilde{p}(\mathbf{x})d\mathbf{x}+\int_{\Omega}
\mathbf{b}(\mathbf{x}).\nabla \mathcal{N}(\mathbf{x},\mathbf{x}_0)
\tilde{p}(\mathbf{x}) d\mathbf{x} =
C_{\epsilon}e^{-\frac{\Phi(\mathbf{x}_0)}{D}}+O\left(1\right).\label{approx}
\eeq
Consequently, (\ref{I2}) reduces to
\beqq
\tilde{p}(\mathbf{x}_0)+O\left(1\right)&=&-\int_{\partial
N_a}
\mathbf{\tilde{J}}(\mathbf{x}).\mathbf{n}_{\mathbf{x}}\mathcal{N}(\mathbf{x},\mathbf{x}_0)
d\mathbf{x}+C_{\epsilon}e^{-\frac{\Phi(\mathbf{x}_0)}{D}}\nonumber\\
\label{I3}
\eeqq
We now compute $\int_{\partial N_a} \mathbf{\tilde{J}}(\mathbf{x}).\mathbf{n}_{\mathbf{x}}\mathcal{N}(\mathbf{x},\mathbf{x}_0)
d\mathbf{x}=\sum_{i=1}^{n} \int_{ \p\Omega_i}
\mathbf{\tilde{J}}(\mathbf{x}).\mathbf{n}_{\mathbf{x}}\mathcal{N}(\mathbf{x},\mathbf{x}_0)
d\mathbf{x}$, by decomposing the flux
\beq
\left(\mathbf{\tilde{J}}(\mathbf{x}).\mathbf{n}_{\mathbf{x}}\right)_{\x
\in \p \Omega_i}=g_i(\x)+f_i(\x),
\label{expansion}
\eeq
where the leading order $g_i(s)$
with
\beq
s=|\mathbf{x}-\mathbf{x}_i|
\eeq
into
\beq
g_i(s)=\frac{g_0^i}{\sqrt{\epsilon^2-s^2}},\label{helmholtz}
\eeq
and  $g_0^i$ a constant and $f_i$ is a regular function such that
\beq
\int_{0}^{\epsilon} f_i(s)  ds=O(\epsilon g_0^i).\label{assumption}
\eeq
Choosing $\x_0=\x_i$ at the absorbing boundary condition, we get
that  $\tilde{p}(\mathbf{x}_i)=0$. For $i\neq j$, and $|\mathbf{x}_i-\mathbf{x}_j| \gg \epsilon$ and
that for $\mathbf{x} \in \p \Omega_j$,
\beq
\mathcal{N}(\mathbf{x},\mathbf{x}_i)= \mathcal{N}(\mathbf{x}_j,\mathbf{x}_i)+O(\epsilon).
\eeq
Consequently, using the flux expansion \ref{expansion}, we get
\beq \label{flux}
\int_{\partial N_a} \mathbf{\tilde{J}}(\mathbf{x}).\mathbf{n}_{\mathbf{x}}\mathcal{N}(\mathbf{x},\mathbf{x}_i)
d\mathbf{x} =\int_{\partial \Omega_i} \left(g_i(\x)+f_i(\x)\right) \mathcal{N}(\mathbf{x},\mathbf{x}_i)
d\mathbf{x}
\eeq
\beqq
+\sum_{j=1,j\neq i}^{n} \left(\mathcal{N}(\mathbf{x_j},\mathbf{x}_i)+0(\epsilon)\right) \int_{\partial \Omega_j} \left(g_j(\x)+f_j(\x)\right) d\mathbf{x}.
\eeqq
For $\mathbf{x}_i$ on the domain boundary $\p S_a$, the Neumann-Green's function $\mathcal{N}(\mathbf{x},\mathbf{x}_i)$
can be written as \cite{leakage}:
\beq
\mathcal{N}(\mathbf{x},\mathbf{x}_i)=\frac{1}{2\pi D
|\mathbf{x}-\mathbf{x}_i|}+\frac{L(\x_i)+N(\x_i)}{8\pi D} \log
\left(\frac{1}{|\x-\x_i|}\right)+\omega_{\mathbf{x}_i}(\mathbf{x}),\label{neumann}
\eeq
where $L(\x_i)$ and $N(\x_i)$ are the principal curvatures of $\p
S_a$ at $\x_i$ and $\omega_{\mathbf{x}_i}(\x)$ is the regular part of the Green function, which is bounded for $\x$ in $\Omega$.

However, when the absorbing small patches are located on the boundary of a small ball of radius $a$, The Green-Neurmann's expansion \ref{neumann} of $\mathcal{N}(\x_i,\x_j)$ does not hold,  because the second term $\frac{-1}{4\pi a D} \log\left(\frac{1}{|\x_i-\x_j|}\right)$ can become much larger than the first term $\frac{1}{2\pi D
|\x_i-\x_j|}$ when $|\x_i-\x_j|\approx a$, and $a\ll |\Omega|^{\frac{1}{3}}$.

\subsection{Analysis for a small internal ball}
We start with the revisited solution of the Neumann's equation
\beqq
D\Delta \mathcal{\tilde{N}}(\x,\x_0)&=&-\delta(\x-\x_0) \hbox{, for  } \x \in \mathbb{R}^3 \nonumber \\
D \frac{\p \mathcal{\tilde{N}}}{\p n} (\x,\x_0) &=& 0 \hbox{, for } \x \in S_a.
\eeqq
which is equal for $|\x_0|=|\x|=a$ to (see appendix)
\beq
\ds{\mathcal{\tilde{N}}(\x,\x_0)=\frac{1}{2\pi D|\x-\x_0|}+ \frac{1}{4\pi a D}\log\left(\frac{|\x-\x_0|}{2a+|\x-\x_0|}\right)}. \label{neumann_sphere}
\eeq
Thus for $\x$ and $\x_0$ in the neighborhood of the sphere $S_a$, we have
\beq
\mathcal{N}(\x,\x_0)=\mathcal{\tilde{N}}(\x,\x_0)+O(1) \label{neumdev}.
\eeq
Consequently, using the geodesic distance $s=d(P,\x_i)$, expanding the flux term in relation \ref{flux}
gives
\beqq
\int_{\partial N_a} \mathbf{\tilde{J}}(\mathbf{x}).\mathbf{n}_{\mathbf{x}}\mathcal{N}(\mathbf{x},\mathbf{x}_i)
d\mathbf{x}&=&\int_{0}^{\epsilon} \left(\frac{g_0^i}{\sqrt{\epsilon^2-s^2}}+f_i(s)\right)\\
&& \left(\frac{1}{2\pi Ds}+ \frac{1}{4\pi a D}\log\left(\frac{s}{2a+s}\right)+O(1)\right) 2\pi s ds \nonumber \\
 &+&\sum_{j=1,j\neq i}^{n}\left(\mathcal{N}(\mathbf{x}_j,\mathbf{x}_i)+0(\epsilon)\right) \int_{0}^{\epsilon} \left( \frac{g_0^j}{\sqrt{\epsilon^2-s^2}}+f_j(s)\right) 2\pi s ds \nonumber.
\eeqq
Using condition (\ref{assumption}), we obtain:
\beqq
\int_{\partial N_a} \mathbf{\tilde{J}}(\mathbf{x}).\mathbf{n}_{\mathbf{x}}\mathcal{N}(\mathbf{x},\mathbf{x}_0) d\mathbf{x}&=&\frac{g_0^i}{D}\left(\frac{\pi}{2}+\frac{\epsilon}{2a} \log\left(\frac{\epsilon}{a}\right)+O(\epsilon)\right)\nonumber \\
 &+&2\pi \epsilon \sum_{j=1,j\neq i}^{n} \mathcal{N}(\mathbf{x}_j,\mathbf{x}_i)  g_0^j \left(1+O(\epsilon))\right). \label{calcul}
\eeqq
We recall that the constant $g_0^i$ is of order $\ds g_0^i=O\left(\frac{1}{n\epsilon}\right)$, and that
\beq
\mathcal{N}(\mathbf{x}_j,\mathbf{x}_i)=O\left(\frac{1}{|\x_i-\x_j|}\right)=O\left(\frac{1}{a}\right),
\eeq
thus we rewrite the flux condition as
\beqq
\int_{\partial N_a} \mathbf{\tilde{J}}(\mathbf{x}).\mathbf{n}_{\mathbf{x}}\mathcal{N}(\mathbf{x},\mathbf{x}_0) d\mathbf{x}&=&\frac{g_0^i}{D}\left(\frac{\pi}{2}+\frac{\epsilon}{2a} \log\left(\frac{\epsilon}{a}\right)\right) \\
 &+&2\pi \epsilon \sum_{j=1,j\neq i}^{n} \mathcal{N}(\mathbf{x}_j,\mathbf{x}_i)  g_0^j +O\left(\frac{\epsilon}{a}\right)+O\left(\frac{1}{n}\right)\nonumber. \label{calcul2}
\eeqq
Injecting \ref{calcul2} in \ref{I3}, for $\x_0=\x_i$, we obtain the system of n equations to solve in the n+1 variable $(g_0^1,..g_0^n,C_{\epsilon})$:
\beq
\left(\frac{\pi}{2D}+\frac{\epsilon}{2aD} \log\left(\frac{\epsilon}{a}\right)\right) g_0^i + 2\pi \epsilon
\sum_{j=1,j\neq i}^{n} \mathcal{N}(\mathbf{x}_j,\mathbf{x}_i)
g_0^j =C_{\epsilon}e^{-\frac{\Phi(\mathbf{x}_i)}{D}}+O\left(1\right).\label{I4}
\eeq
To close the system of equation, we use the compatibility condition (eq. \ref{conservation}) with expression \ref{expansion} and approximation \ref{LTAP}  for the function  $\tilde{p}$:
\beq
2\pi \epsilon \sum_{i=1}^{n}g_0^i =1-C_{\epsilon}\int_{\Omega}k(\mathbf{x})e^{-\frac{\Phi(\mathbf{x})}{D}}d\mathbf{x}+O\left(1\right),
\label{conservation2}
\eeq
Finally, we obtain a linear system of $n+1$
equations (\ref{I4}) and (\ref{conservation2}) for the flux constant
$g_0^i$ ($i\leq i \leq n$), and for the parameter $C_{\epsilon}$,
summarized as
\beq\label{linear}
 \left\{\begin{array}{ll}
\ds{\frac{\pi}{2D}+\frac{\epsilon}{2aD} \log\left(\frac{\epsilon}{a}\right) g_0^i+2\pi  \epsilon  \sum_{j=1,j\neq i}^{n}\mathcal{N}(\mathbf{x}_j,\mathbf{x}_i) g_0^j }=\ds{ C_{\epsilon}e^{-\frac{\Phi(\mathbf{x}_i)}{D}}}+O\left(1\right),
\hbox{ for }1\leq i \leq n   \\   \\
\ds{2\pi \epsilon \sum_{i=1}^{n}g_0^i} =
\ds{1-C_{\epsilon}\int_{\Omega}k(\mathbf{x})e^{-\frac{\Phi(\mathbf{x})}{D}}d\mathbf{x} +O\left(1\right)}
\end{array}\right.
\eeq
We will now obtain asymptotic estimates for $C_{\epsilon}$, $\langle P \rangle $ and $\langle \tau \rangle$, by solving the linear system of equations \ref{linear} in the limit $\epsilon$ small. Injecting expressions \ref{LTAP} and \ref{LTAQ} for $\tilde{p}(\x)$ and $q(\x)$ in \ref{PN}-\ref{TN}, we obtain
\beq
\langle P \rangle =1-C_{\epsilon}\int_{\Omega}k(\mathbf{x})e^{-\frac{\Phi(\mathbf{x})}{D}}d\mathbf{x}
\hbox{ and }
\langle \tau \rangle=C_{\epsilon}\int_{\Omega}e^{-\frac{\Phi(\mathbf{x})}{D}}d\mathbf{x}.
 \label{proba}
 \eeq
We derive the asymptotic expression in the next section.
\section{Mean field approximation and asymptotics formula for  $\langle \tau \rangle$ and $\langle P \rangle$ for $n \gg \frac{1}{\epsilon}$}
We derive now expressions for $\langle P \rangle$ and $\langle \tau \rangle$ in the limit $n\gg1$ and absorbing windows are distributed with a density $\rho(\x)$ over the spherical nucleus $S_a$. By summing equations eq. \ref{I4} for $1 \leq i \leq n$, we obtain that
\beq
\left(\frac{\pi}{2D}+\frac{\epsilon}{2aD} \log\left(\frac{\epsilon}{a}\right)\right) \sum_{i=1}^n g_0^i + 2\pi \epsilon
\sum_{i=1}^n g_0^i \sum_{j=1,j\neq i}^{n} \mathcal{N}(\mathbf{x}_j,\mathbf{x}_i)
 = C_{\epsilon} \sum_{i=1}^n e^{-\frac{\Phi(\mathbf{x}_i)}{D}}+O\left(n\right).  \label{hj}
\eeq
When $\x_i$ is located at the north pole, the distance $|\x_i-\x_j|$ with $j^{th}$ located at position $\x_j(\theta,\phi)$ is given by $|\x_i-\x_j|=2 a \sin\left(\frac{\phi}{2}\right)$ and the Neumann function is
\beq
\mathcal{\tilde{N}}(\x_j(\theta,\phi),\x_i)=\frac{1}{4\pi a D}\left(\frac{1}{\sin\left(\frac{\phi}{2}\right)}+\log\left(\frac{\sin\left(\frac{\phi}{2}\right)}{1+ \sin\left(\frac{\phi}{2}\right)}\right)\right). \label{fi}
\eeq
We use now that the probability density function $\rho_i(\phi)$ of the $j\neq i $ windows (north pole $i$) is normalized  by the condition
\beq
\int_0^{\pi} 2\pi a^2 \rho_i(\phi) \sin(\phi)d\phi  = 1,
\eeq
thus
\beqq
\lim_{n \to \infty} \frac{1}{n}\sum_{j=1,j\neq i}^{n} \mathcal{N}(\mathbf{x}_j,\mathbf{x}_i) &=&  \int_0^{\pi} \frac{1}{4\pi a D} \left(\frac{1}{\sin\left(\frac{\phi}{2}\right)}+\log\left(\frac{\sin\left(\frac{\phi}{2}\right)}{1+ \sin\left(\frac{\phi}{2}\right)}\right)\right) \rho_i(\phi) 2 \pi a^2 \sin(\phi)d\phi \nonumber \\
&=& \frac{ a}{2D} \int_{0}^{\pi} \left(\frac{1}{\sin\left(\frac{\phi}{2}\right)} +\log\left(\frac{\sin\left(\frac{\phi}{2}\right)}{1+ \sin\left(\frac{\phi}{2}\right)}\right)\right) \rho_i(\phi) \sin(\phi)d\phi.
\eeqq
In addition, we can also approximate
\beq
\lim_{n \to \infty} \frac{1}{n}\sum_{i=1}^{n} e^{-\frac{\Phi(\mathbf{x}_i)}{D}} = a^2 \int_{0}^{2\pi} \int_{0}^{\pi} e^{-\frac{\Phi(\phi,\theta)}{D}} \rho(\phi)d\phi d\theta, \label{int2}
\eeq

We now re-write relation \ref{hj} using
\beq
I_1^i = \int_{0}^{\pi} \left(\frac{1}{\sin\left(\frac{\phi}{2}\right)}+\log\left(\frac{\sin\left(\frac{\phi}{2}\right)}{1+ \sin\left(\frac{\phi}{2}\right)}\right)\right) \rho_i(\phi) \sin(\phi)d\phi,
\eeq
\beq
I_2 = \int_0^{2\pi}\int_{0}^{\pi} e^{-\frac{\Phi(\phi,\theta)}{D}} \rho(\phi,\theta)\sin(\phi)d\phi d\theta,
\eeq
so that
\beq
\left(\frac{\pi}{2D}+\frac{\epsilon}{2aD} \log\left(\frac{\epsilon}{a}\right)\right) \sum_{i=1}^n g_0^i + \frac{na \pi \epsilon}{D}
\sum_{i=1}^n g_0^i I_1^i
 = C_{\epsilon} n a^2 I_2+O\left(n\right).  \label{hi2}
\eeq
For identically distributed windows $I_1^i = I_1$ with $1\leq i \leq n$. Using the compatibility condition \ref{conservation2} in equation \ref{hi2}, we obtain
\beq
\left(\frac{1}{4D\epsilon}+\frac{1}{4\pi aD} \log\left(\frac{\epsilon}{a}\right) + \frac{na }{2D} I_1\right)
\left(1-C_{\epsilon}\int_{\Omega}k(\mathbf{x})e^{-\frac{\Phi(\mathbf{x})}{D}}d\mathbf{x}\right)
 = C_{\epsilon} n a^2 I_2+O\left(n\right).
\eeq
Thus we obtain to leading order
\beq
C_{\epsilon}=\frac{\pi a+\epsilon \log\left(\frac{\epsilon}{a}\right) + 2n\pi a^2 \epsilon  I_1}
{\left(\pi a+\epsilon \log\left(\frac{\epsilon}{a}\right) + 2n\pi a^2 \epsilon  I_1\right)\int_{\Omega}k(\mathbf{x})e^{-\frac{\Phi(\mathbf{x})}{D}}d\mathbf{x}+ 4 \pi n a^3 D \epsilon I_2} \label{Ce}
\eeq
To further compute for the probability $\langle P \rangle$ and the MFPT $\langle \tau \rangle$ using expression \ref{Ce}, we shall consider two distributions of windows:
\begin{enumerate}
\item Random distribution
\item Uniform distribution
\end{enumerate}
\subsection{Random distribution of narrow windows located on a sphere}
When there $n\gg 1$ non-overlapping windows randomly distributed on the sphere, the probability distribution of windows is given by
\beq
\rho(\phi,\theta)=\rho_i(\phi)= \frac{1}{4\pi a^2} \mathbf{1}_{ \left\{\phi > 2 \arcsin\left(\frac{\epsilon}{a}\right)\right\}},
\eeq
for all $1\leq i \leq n$. The condition $\left\{\phi > 2 \arcsin\left(\frac{\epsilon}{a}\right)\right\}$ ensures non-overlapping.
Changing the variable $y = \sin\left(\frac{\phi}{2}\right)$, we re-write
\beq
I_1 = \frac{1}{\pi a^2} \int_{\frac{\epsilon}{a}}^{1} \left(\frac{1}{y}+\log\left(\frac{y}{1+ y}\right)\right)  y dy= \frac{1}{2\pi a^2} \left[x+\log\left(1+x\right)+x^2\log\left(\frac{x}{1+x}\right)\right]_{\frac{\epsilon}{a}}^1,
\eeq
that is
\beq
I_1  = \frac{1}{2\pi a^2} \left(1-2\frac{\epsilon}{a}-\frac{\epsilon^2}{a^2}\log\left(\frac{\epsilon}{a}\right)\right). \label{I1_rand}
\eeq
In addition, we have
\beq
I_2  = \frac{1}{4\pi a^2} \int_0^{2\pi}\int_{2 \arcsin\left(\frac{\epsilon}{a}\right)}^{\pi} e^{-\frac{\Phi(\phi,\theta)}{D}} \sin(\phi)d\phi d\theta. \label{I2_rand}
\eeq
Replacing in eq. \ref{Ce}, $I_1$ and $I_2$ by expressions \ref{I1_rand} and \ref{I2_rand}, we obtain to leading order for randomly distributed windows,

\beq
C_{\epsilon}^{\text{rand}}=\frac{1}
{\ds \int_{\Omega}k(\mathbf{x})e^{-\frac{\Phi(\mathbf{x})}{D}}d\mathbf{x}+  C(n,\eps) \int_0^{2\pi}\int_{2\frac{\epsilon}{a}}^{\pi} e^{-\frac{\Phi(\phi,\theta)}{D}} \sin(\phi)d\phi d\theta}. \label{Ce_rand}
\eeq
where
\beq
C(n,\eps)= \frac{n a D \epsilon}{\pi a+\epsilon\left(1-\frac{n\epsilon^2}{a^2}\right) \log\left(\frac{\epsilon}{a}\right) + n \epsilon  \left(1-2\frac{\epsilon}{a}\right)}.
\eeq
\subsection{Homogeneous distribution of windows on the surface $S_a$}
For small windows homogeneously distributed on a sphere, the density is given by \cite{cheviakov2013}
\beq
\rho(\phi) = \mathbf{1}_{\left\{\phi> \arccos\left(1-\frac{2}{n}\right) \right\} \frac{1}{4\pi a^2}}
\eeq
leading to
\beq
I_1  = \frac{1}{2\pi a^2} \left[x+\log\left(1+x\right)+x^2\log\left(\frac{x}{1+x}\right)\right]_{\frac{1}{2}\arccos\left(1-\frac{2}{n}\right)}^1,
\eeq
and for $n\gg1$
\beq \label{I1_hom}
I_1  = \frac{1}{2\pi a^2} \left(1-\frac{2}{\sqrt{n}}+\frac{\log(n)}{2n}\right) +o(\frac{\log(n)}{2n}).
\eeq
In addition,  we have
\beq
I_2  = \frac{1}{4\pi a^2} \int_0^{2\pi}\int_{2 \arccos\left(1-\frac{2}{n}\right)}^{\pi} e^{-\frac{\Phi(\phi,\theta)}{D}} \sin(\phi)d\phi d\theta. \label{I2_hom}
\eeq
Replacing in equation \ref{Ce}, $I_1$ and $I_2$ by expressions \ref{I1_hom} and \ref{I2_hom} respectively, leads to
\beq
C_{\epsilon}^{\text{hom}}=\frac{1}
{\int_{\Omega}k(\mathbf{x})e^{-\frac{\Phi(\mathbf{x})}{D}}d\mathbf{x}+  \tilde C_{\epsilon,n} \int_0^{2\pi}\int_{\frac{4}{\sqrt{n}}}^{\pi} e^{-\frac{\Phi(\phi,\theta)}{D}} \sin(\phi)d\phi d\theta}. \label{Ce_hom}
\eeq
where
\beq
\tilde C_{\epsilon,n}=\frac{n a D \epsilon }{
\pi a+\epsilon \log\left(\frac{\sqrt{n}\epsilon}{a}\right) + n \epsilon  \left(1-\frac{2}{\sqrt{n}}\right)}
\eeq
Using formula \ref{Ce_rand} and \ref{Ce_hom} in expression \ref{proba}, we obtain the asymptotic expressions for the probability and the condition MFPT that a stochastic particle reaches a small windows
\beq
\langle P \rangle =
\frac{F(n,a,\epsilon) \int_0^{2\pi}\int_{\alpha_0}^{\pi} e^{-\frac{\Phi(\phi,\theta)}{D}} \sin(\phi)d\phi d\theta}{\int_{\Omega}k(\mathbf{x})e^{-\frac{\Phi(\mathbf{x})}{D}}d\mathbf{x}+ F(n,a,\epsilon) \int_0^{2\pi}\int_{\alpha_0}^{\pi} e^{-\frac{\Phi(\phi,\theta)}{D}} \sin(\phi)d\phi d\theta}, \label{proba_final}
\eeq
%
and
\beq
\langle \tau \rangle = \ds \frac{ \int_{\Omega}e^{-\frac{\phi(\mathbf{x})}{D}}d\mathbf{x}}
{\int_{\Omega}k(\mathbf{x})e^{-\frac{\Phi(\mathbf{x})}{D}}d\mathbf{x}+ F(n,a,\epsilon)\int_0^{2\pi}\int_{\alpha_0}^{\pi} e^{-\frac{\Phi(\phi,\theta)}{D}} \sin(\phi)d\phi d\theta} \label{MFPT_final}
\eeq
where
\beq
F(n,a,\epsilon)=\frac{n a D \epsilon}{\left(\pi a+\epsilon \log\left(\frac{\epsilon}{a}\right) + n\epsilon\left(1-2\alpha_0-\alpha_0^2\log\left(\alpha_0\right)\right) \right)}
\eeq
and
\beq
\alpha_0= \left\{\begin{array}{l} \ds \frac{\epsilon}{a} \quad\mbox{ for uniformly randomly distributed windows,}\\ \\
\ds \frac{1}{\sqrt{n}}\quad \quad\mbox{ for homogeneously distributed windows},
\end{array}\right.\label{eq12}
\eeq
When the drift is pointing towards the nucleus center and the potential $\Phi(\x)=\Phi_0$ is constant at nuclear surface, then the probability and MFPT formulas reduce to
\beq
\langle P \rangle = \frac{ 4\pi F(n,a,\epsilon) e^{-\frac{\Phi_0}{D}}}{\int_{\Omega}k(\mathbf{x})e^{-\frac{\phi(\mathbf{x})}{D}}d\mathbf{x}+ 4\pi F(n,a,\epsilon)e^{-\frac{\Phi_0}{D}} }, \label{proba_final_ct}
\eeq
%
and
\beq
\langle \tau \rangle = \frac{\int_{\Omega}e^{-\frac{\phi(\mathbf{x})}{D}}d\mathbf{x}}{\int_{\Omega}k(\mathbf{x})e^{-\frac{\phi(\mathbf{x})}{D}}d\mathbf{x}+ 4\pi F(n,a,\epsilon) e^{-\frac{\Phi_0}{D}} }. \label{MFPT_final_ct}
\eeq
When the drift $\Phi_{S_a}$  restricted to $S_a$ has a single global minima $\Phi_m$ at position $\mathbf{x}_0 (\phi_0,\theta_0) \in S_a$,  we approximate integral $\ds{I_2}$ using Laplace's method. In the small diffusion limit $D\ll \Phi(\mathbf{x})$ and large n, we get
\beq
I_2= \frac{1}{4\pi a^2} \int_0^{2\pi}\int_{2 \arccos\left(1-\frac{2}{n}\right)}^{\pi} e^{-\frac{\Phi(\phi,\theta)}{D}} \sin(\phi)d\phi d\theta \approx \frac{D}{4 a^2 \sqrt{\det\left[-H_{\Phi_{S_a}}(\mathbf{x}_0)\right]}} e^{-\frac{\Phi_m}{D}} \label{I2approx}
\eeq
where $\det\left[H_{\Phi_{S_a}}(\mathbf{x}_0)\right]$ is the determinant of the Hessian matrix of potential $\Phi_{S_a}$ at $\mathbf{x}_0$. The probability and MFPT to a nuclear pore are then given by
\beq
\langle P \rangle =
\frac{\pi D F(n,a,\epsilon) \sqrt{\det^{-1}\left[H_{\Phi_{S_a}}(\mathbf{x}_0)\right]} e^{-\frac{\Phi_m}{D}}}{ \int_{\Omega}k(\mathbf{x})e^{-\frac{\Phi(\mathbf{x})}{D}}d\mathbf{x}+ \pi D F(n,a,\epsilon) \sqrt{\det^{-1}\left[H_{\Phi_{S_a}}(\mathbf{x}_0)\right]} e^{-\frac{\Phi_m}{D}}}, \label{proba_final_hess}
\eeq
and
\beq
\langle \tau \rangle = \frac{ \int_{\Omega}e^{-\frac{\phi(\mathbf{x})}{D}}d\mathbf{x}}
{\int_{\Omega}k(\mathbf{x})e^{-\frac{\Phi(\mathbf{x})}{D}}d\mathbf{x}+ \pi D F(n,a,\epsilon) \sqrt{\det^{-1}\left[H_{\Phi_{S_a}}(\mathbf{x}_0)\right]} e^{-\frac{\Phi_m}{D}}} \label{MFPT_final_hess}
\eeq
A second Laplace's method can be used to estimate the volume integral. If the global minimum $\Phi_{\Omega}$ is attained at a point $\x_{g} \in \Omega$,
\beq
\ds \int_{\Omega}e^{-\frac{\phi(\mathbf{x})}{D}}d\mathbf{x}\approx  \frac{(\pi D)^{3/2}}{\sqrt{\det\left[H_{\Phi}(\mathbf{x}_g)\right]}} e^{-\frac{\Phi_{\Omega}}{D}}.
\eeq
We conclude this section by indicating that the formulas presented above can be used to estimate the probability and the mean time for a stochastic viral particle to reach a nuclear pore inside the nucleus.

\subsection{Effect of changing the window coverage on the escape time }
For a large  windows $n\gg1$, distributed over a small surface $S_a$ of a domain $\Omega$, the leading order term of the narrow escape time for a Brownian particle to one of the small window was derived using electrostatic \cite{juergen}
\beq
\langle \tau\rangle_{ES} =\frac{|\Omega|}{D}\left(\frac{1}{C_{S_a}}+\frac{f(\sigma)}{4
n \epsilon}\right),\label{tau-st}
\eeq
where $|\Omega|$ is the volume, $C_{S_a}$ is the capacity of the surface $\partial S_a$ where absorbing holes are distributed, and
\beq
\sigma=\frac{N\pi \epsilon^2}{|\partial S_a|}
\eeq
is the fraction of $S_a$ covered by the absorbing holes. When the surface $S_a$ is a sphere of radius $a$, then $C_{S_a}=4\pi a$, and the MFPT is given by
\beq
\langle \tau\rangle_{ES} =\frac{|\Omega|}{D}\left(\frac{1}{4\pi a}+\frac{f(\sigma)}{4
n \epsilon}\right).
\eeq
In general, the function $f(\sigma)$ is unknown, but is given to leading order by $f(\sigma)=1$ \cite{Berg-Purcell}. Here, for a Brownian particle (no drift and no killing measure), the MFPT (eq. \ref{MFPT_final}) reduces to
\beq
\langle \tau\rangle_{\Phi=0,k=0} \approx \ds{\frac{|\Omega|}{D}\left(\frac{1}{4\pi a} + \frac{1}{4 n\epsilon} \left(1-\frac{n\epsilon}{\pi a} \left(2\alpha_0-\alpha_0^2\log(\alpha_0)+\frac{1}{n}\log\left(\frac{\epsilon}{a}\right) \right)\right)\right)}
\eeq

Thus, we identify here the function
\beq
f(\sigma)=1-8\frac{\sigma}{\pi}+\frac{\epsilon}{a\pi}(1-4\sigma)\log\left(\frac{\epsilon}{a}\right)+o\left(\frac{\epsilon}{a}\right) \label{fsigma_rand}
\eeq
when non-overlapping absorbing holes are randomly distributed and
\beq
f(\sigma)=1-4\frac{\sqrt{\sigma}}{\pi}+\frac{\epsilon}{a\pi}\log\left(\sqrt{\sigma}\right)+o\left(\frac{\epsilon}{a}\right) \label{fsigma_hom}
\eeq
when absorbing holes are distributed homogeneously.

We end this section with two remarks. First, for $\sigma<<1$, $8\frac{\sigma}{\pi}<4\frac{\sqrt{\sigma}}{\pi}$, the MFPT of a single particle to an absorbing hole is higher for randomly distributed holes compared to homogeneously distributed holes. Second, formulas \ref{fsigma_rand} and \ref{fsigma_hom} derived here by accounting for two window coverage predict MFPT formula different then previously reported based on an effective medium treatment ($f(\sigma)=1-\sigma$ \cite{zwanzig}) or interpolated from Brownian simulations ($f(\sigma)=\frac{\ds{1-\sigma}}{\ds{1+3.8 \sigma^{1.25}}}$ \cite{berezhkovskii}). This  difference may arise from the differences in the window arrangements.

\section{Comparison of asymptotic formula with respect to Brownian simulations}
\begin{figure}
\includegraphics[width=8cm]{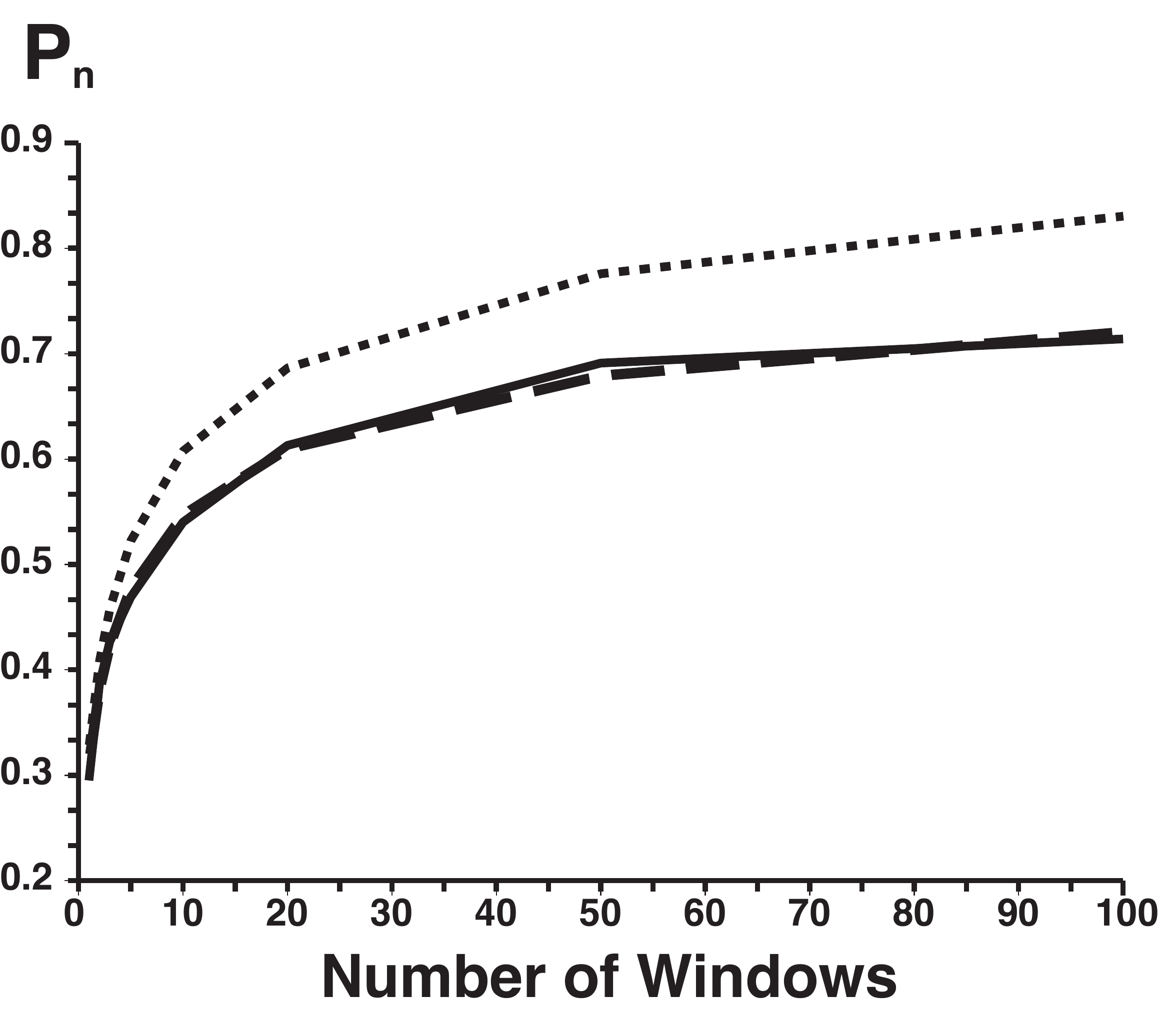}
\includegraphics[width=8cm]{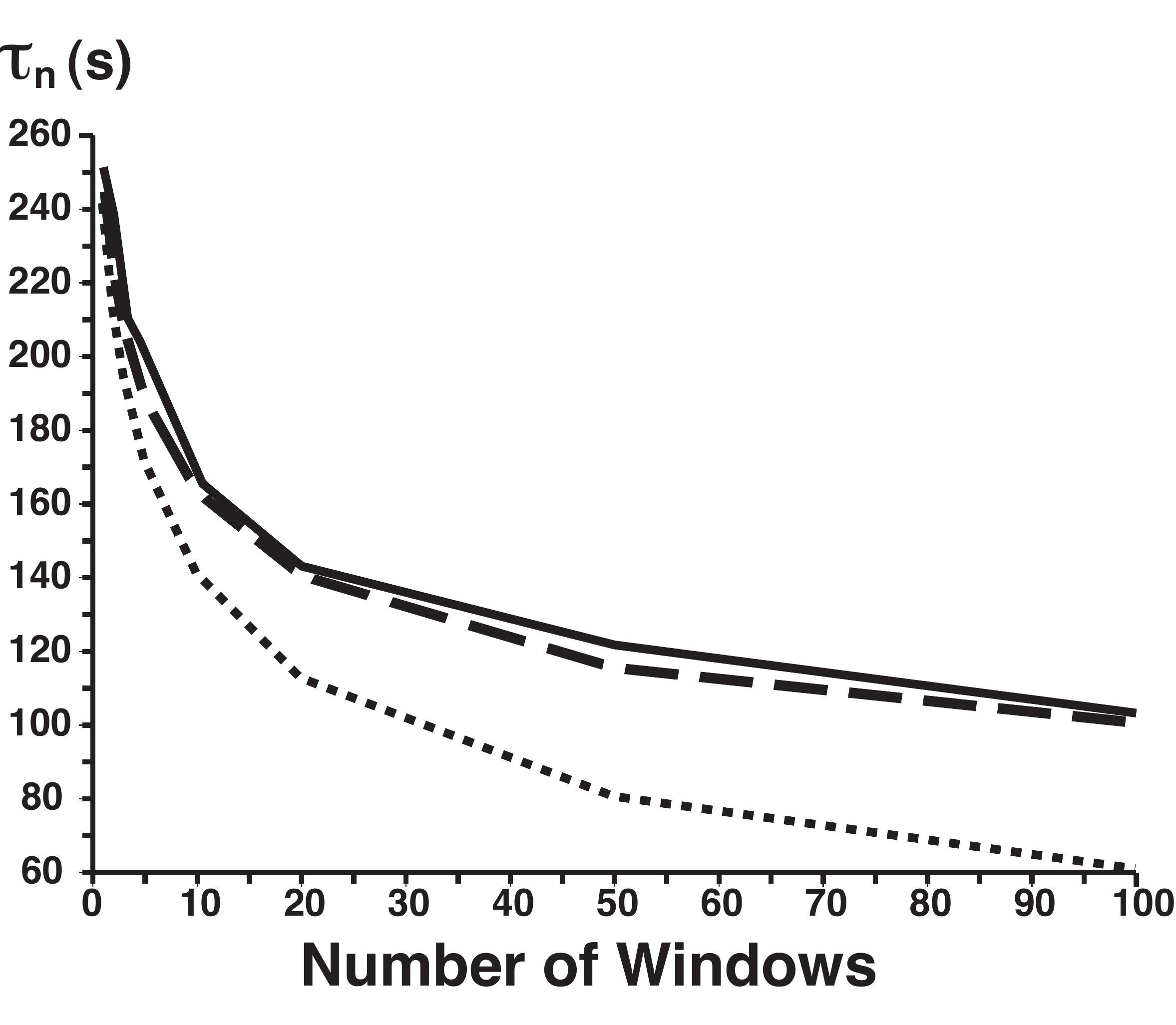}
\caption{\small{\bf Probability $P_n$ (left) and conditional MFPT $\tau_n$ (right).}
Asymptotics formula \ref{proba3} and \ref{temps3} (dashed line) are compared to stochastic simulations (solid line) for an increasing number of absorbing windows, while the ratio $\sigma=\frac{n\pi \epsilon^2}{4\pi a^2}$ of the absorbing to the total surface is kept constant $\sigma=2\%$ \cite{maul}. The asymptotics formula \ref{PN} that do not account for the window interactions is presented to visualize the improvement of the new formula (dotted line). $1000$ stochastic trajectories are simulated. Parameters are summarized in table \ref{par}.}\label{FIGURE2}
\end{figure}

For a ball of radius $R$ with a centered sphere $S_a$ (radius $a$) uniformly covered by $n$ small absorbing pores (radius $\epsilon$) (Fig. \ref{FIGURE1} right). Stochastic particles are reflected on the external membrane $r=R$ and on $r=a$ except on all windows $\p N_a=\bigcup_{i=1}^{n} \p \Omega_i$, centered at random locations $\left(\mathbf{x}_i\right)_{i=1}^{n}$. We use a constant radial drift $B$ directed toward the nucleus (with a potential $\Phi(r)=-Br$). We consider a constant killing rate $k(\mathbf{x})=k_0$ and consequently, using function
\beq
G(D,B,a)=e^{-\frac{Ba}{D}}\left(\frac{D}{B}a^2+2\left(\frac{D}{B}\right)^2a +2\left(\frac{D}{B}\right)^3\right)
\eeq
expressions (\ref{proba_final_ct}-\ref{MFPT_final_ct}) simplify to
\beq
\langle P \rangle = \frac{e^{-\frac{Ba}{D}}}{\langle \tau\rangle_{\Phi=0,k=0} \left(G(D,B,a)-G(D,B,R)\right)k+e^{-\frac{Ba}{D}}}, \label{proba3}
\eeq
and
\beq
\langle \tau\rangle =
\frac{\langle \tau\rangle_{\Phi=0,k=0} \left(G(D,B,a)-G(D,B,R)\right)}{\langle \tau\rangle_{\Phi=0,k=0}\left(G(D,B,a)-G(D,B,R)\right)k+e^{-\frac{Ba}{D}}}
\label{temps3}.
\eeq
In Fig. \ref{FIGURE2}, we show how these expressions compare to stochastic simulations for an increasing number of holes while maintaining  constant the ratio $\sigma=\frac{n \pi \epsilon^2}{4\pi a^2}$ of the nucleus surface covered by the absorbing windows to the value $\sigma=2\%$. This number was calibrated by using a surface covered by $2,000$ pores of $25nm$ diameter on the nucleus of a chinese hamster ovary cell \cite{maul}). The parameters are summarized in table \ref{par}.

\section{Conclusion}
Intermittent dynamics with alternative periods of free diffusion and directed motion along MTs characterizes a large class of cellular transports. When the intermittent particle can be degraded through the ubiquitin-proteasome machinery or trapped in the crowded cytoplasm, we derived here improved asymptotics formula for the probability $P_n$ and the mean time $\tau_n$ to reach a small absorbing target among $n$. These formula accounts for the geometrical interactions between the windows. When the targets co-localize on a small domain $S_a$, asymptotics of $P_n$ and $\tau_n$ are obtained in the limit $\frac{|S_a|}{|\Omega|}\ll 1$. Applied to DNA viruses that have to reach a small nuclear pore among the $2,000$, these formulas provide estimates for the arrival time to the nucleus. We confirmed here the validity of asymptotics formula \ref{proba3}-\ref{temps3}) for the probability $P_n$ and the mean time $\tau_n$ respectively using Brownian simulations. We note that contrary to the classical narrow escape asymptotic where the leader order term contains most of the  geometry, here the $O(1)-$term accounts for the interactions between windows. For example, where there are $100$ windows, the asymptotic formula of the conditioned MFPT gives $\tau_n\approx 2 min$, similar to simulation results, but is twice the one derived in \cite{david}, for which $\tau_n\approx1 min.$. 


\begin{center}
\begin{table}[http!]
\begin{tabular}{c|l|l}
\hline
\hline
Parameter  & Description & Value \\
 \hline
$D$ & Diffusion constant of the virus & $D=1.3\mu m^2s^{-1}$ (Observed \\
      &&for the Associated-Adeno-Virus \cite{Seisenberger}) \\
      $B$ & Drift & $B=0.2 \mu ms^{-1}$  \cite{PRE-manu} \\
      $\sigma$ & \% of the nuclear surface covered  & $\sigma=2\%$ \cite{maul} \\
      & by $n$ nuclear pores \\
      $k$ & Degradation rate & $k=1/360 s^{-1}$ ($10$ times the rate observed \\
      && for gene vectors \cite{Lechardeur}) \\
      $R$ & Radius of the cell & $R=15\mu m$ (Chinese hamster ovary cell)\\
      $a$ & Radius of the nucleus &$a=5\mu m$\cite{maul} \\
 \hline
\end{tabular}
\caption{Numerical parameters used for Brownian simulations} \label{par}
\end{table}
\end{center}

\section{Appendix}
We derive in this appendix the asymptotic of the Neumann's function $\mathcal{N}(\x_i,\x_j)$ for two absorbing patches localized on the surface of a small ball of radius $a$. In that case, expansion of eq. \ref{neumann} is not sufficient and the parameter $a$ should be now accounted for.

Indeed, the log-term $\frac{-1}{4\pi a D} \log\left(\frac{1}{|\x_i-\x_j|}\right)$ can be much larger than the leading order term $\frac{1}{2\pi D |\x_i-\x_j|}$ when $|\x_i-\x_j|\approx a$, for $a\ll |\Omega|^{\frac{1}{3}}$. Consequently, we shall re-examine the $\log$-term expansion. For $\x$ and $\x_0$ in the neighborhood of the sphere $S_a$, we expand the Neumann function $\mathcal{N}(\x,\x_0)$ as
\beq
\mathcal{N}(\x,\x_0)=\mathcal{\tilde{N}}(\x,\x_0)+O(1),
\eeq
where
$\mathcal{\tilde{N}}(\x,\x_0)$ is solution of with $D=1$,
\beqq
\Delta \mathcal{\tilde{N}}(\x,\x_0)&=&-\delta(\x-\x_0) \hbox{ for } \x \in \mathbb{R}^3 \nonumber \\
\frac{\p \mathcal{\tilde{N}}}{\p n} (\x,\x_0) &=& 0 \hbox{ for } \x \in S_a.
\eeqq
To compute the $\log$-term, we first decompose $\mathcal{\tilde{N}}(\x,\x_0)=\frac{1}{4\pi |\x-\x_0|}+\Phi(\x,\x_0)$ where $\Phi$ is solution of the system:
\beqq
\Delta \Phi(\x,\x_0)&=&0, \hbox{ for } \x \in \mathbb{R}^3 \nonumber \\
\frac{\p \Phi}{\p n} (\x,\x_0) &=& -\frac{\p}{\p n}\left(\frac{1}{4\pi  |\x-\x_0|}\right) \hbox{ ,for } \x \in S_a. \label{equadiff}
\eeqq
To solve eq. \ref{equadiff}, we choose a coordinate system so that the source point $\x=\x_0$ is on the positive $z$ axis. Since $a \Phi=0$ and $\Phi$ is axisymmetric, then $\Phi$  has the series expansion
\beq
\Phi(\x,\x_0)=\sum_{n=0}^{\infty} b_n(|\x_0|) \frac{P_n(\cos(\theta))}{|\x|^{n+1}},
\eeq
where $P_n$ are the Legendre polynomials of integer $n$, $\theta$ is the angle between $\x$ and the north pole and $b_n(|\x_0|)$ are coefficients, determined from boundary condition \ref{equadiff}.

For $\x \in S_a$ and $\rho=|\x|$,
\beq
\frac{\p \Phi}{\p n} (\x,\x_0)=\frac{\p \Phi}{\p \rho} (\rho=a)=-\sum_{n=0}^{\infty}\frac{\left(n+1\right) b_n(|\x_0|)}{a^{n+2}} P_n(\cos(\theta).\label{bound1}
\eeq
On the other hand, for $|\x|<|\x_0|$ we have the expansion
\beq
\frac{1}{4\pi |\x-\x_0|}=\frac{1}{4\pi }\sum_{n=0}^{\infty}\frac{|\x|^n}{|\x_0|^{n+1}}P_n\left(\cos \left(\theta\right)\right),\label{gen}
\eeq
which leads to the boundary condition:
\beq
-\frac{\p}{\p \rho}\left(\frac{1}{4\pi  |\x-\x_0|}\right)\left(\rho=a\right)=-\frac{1}{4\pi }\sum_{n=0}^{\infty}\frac{n a^{n-1}}{|\x_0|^{n+1}}P_n\left(\cos \left(\theta\right)\right).\label{bound2}
\eeq
Injecting relation \ref{bound1}-\ref{bound2}) into the boundary condition \ref{equadiff}, we obtain that for all $n \geq 0$:
\beq \label{b_n}
 b_n(|\x_0|)=\frac{1}{4\pi } \frac{n a^{2n+1}}{(n+1)|\x_0|^{n+1}}.
\eeq
The Neumann function $\mathcal{\tilde{N}}(\x,\x_0)$ is then given by:
\beq
\mathcal{\tilde{N}}(\x,\x_0)=\frac{1}{4\pi |\x-\x_0|}+\frac{1}{4\pi } \sum_{n=0}^{\infty} \frac{n a^{2n+1}}{(n+1)|\x|^{n+1}|\x_0|^{n+1}} P_n(\cos(\theta)),
\eeq
that we rewrite
\beq\label{expa}
\mathcal{\tilde{N}}(\x,\x_0)=\frac{1}{4\pi |\x-\x_0|}+\frac{1}{4\pi } \sum_{n=0}^{\infty} \left(\frac{a^{2n+1}}{|\x|^{n+1}|\x_0|^{n+1}}-\frac{a^{2n+1}}{(n+1) |\x|^{n+1}|\x_0|^{n+1}}\right) P_n(\cos(\theta)).
\eeq
Using expansion \ref{gen}), we have for the first term of \ref{expa}
\beq
\frac{1}{4\pi } \sum_{n=0}^{\infty} \frac{a^{2n+1}}{|\x|^{n+1}|\x_0|^{n+1}} P_n(\cos(\theta))=\frac{a}{4\pi |\x_0||x-\frac{a^2 \x_0}{|\x_0|^2}| }
\eeq
To compute the second term $I(\rho)=-\sum_{n=0}^{\infty}\frac{a^{2n+1}}{ (n+1) \rho^{n+1}|\x_0|^{n+1}} P_n(\cos(\theta))$, we note that
\beq
I'(\rho)=\sum_{n=0}^{\infty}\frac{a^{2n+1}}{ \rho^{n+2}|\x_0|^{n+1}} P_n(\cos(\theta))=\frac{a}{\rho |\x_0||x-\frac{a^2 \x_0}{|\x_0|^2}|},
\eeq
that is
\beq
I'(\rho)=\frac{1}{\rho a \left(1+\frac{|\x_0|^2 \rho^2}{a^4}-2\frac{|\x_0|\rho}{a^2} \cos(\theta)\right)^{\frac{1}{2}}}.
\eeq
Because $\lim_{\rho \to \infty} l(\rho)=0$, we have:
\beq
l(\rho)=-\int_{\rho}^{\infty} I'(s)ds= -\int_{\rho}^{\infty} \frac{ds}{s a \left(1+\frac{|\x_0|^2 s^2}{a^4}-2\frac{|\x_0| s}{a^2} \cos(\theta)\right)^{\frac{1}{2}}}.
\eeq
Thus,
\beq
\ds{l(\rho)=\frac{1}{a}\log\left(\frac{\frac{|\x_0|\rho}{a^2}\left(1-\cos(\theta)\right)}{1-\frac{|\x_0|\rho}{a^2} \cos(\theta)+\left(1+\left(\frac{|\x_0|\rho}{a^2}\right)^2-2\frac{|\x_0|\rho}{a^2} \cos(\theta)\right)^{\frac{1}{2}}}\right)}. \label{I5}
\eeq
Finally, we obtain the expression of the Neumann function $\mathcal{\tilde{N}}(\x,\x_0)$ and the exact dependency with the inner ball radius:
\beqq
\mathcal{\tilde{N}}(\x,\x_0)&=&\frac{1}{4\pi |\x-\x_0|}+\frac{a}{4\pi D |\x_0||x-\frac{a^2 \x_0}{|\x_0|^2}|}\nonumber \\&+&
\ds{\frac{1}{4\pi a }\log\left(\frac{\frac{|\x_0||\x|}{a^2}\left(1-\cos(\theta)\right)}{1-\frac{|\x_0||\x|}{a^2} \cos(\theta)+\left(1+\left(\frac{|\x_0||\x|}{a^2}\right)^2-2\frac{|\x_0||\x|}{a^2} \cos(\theta)\right)^{\frac{1}{2}}}\right)}. \label{N}
\eeqq
When $\x$ and $\x_0$ are on the sphere $S_a$, $|\x_0|=|\x|=a$, we have
\beqq
\mathcal{\tilde{N}}(\x,\x_0)&=&\frac{1}{2\pi |\x-\x_0|}+ \frac{1}{4\pi a }\log\left(\frac{|\x-\x_0|}{2a+|\x-\x_0|}\right). \label{fi2}
\eeqq

{\bf Acknowledgments:} T.L. is supported by a FRM post-doctoral fellowship and a grant from Philippe Foundation. D. H. research is supported by a Marie-Curie fellowship.

\bibliographystyle{spmpsci}
\bibliography{Reference_p1}

\end{document}